\documentclass{elsart}
\usepackage{epsfig}
\usepackage{amssymb}

\begin{document}
\begin{frontmatter}

\title{Possible Observation of Photon Speed Energy Dependence}

\author{V.~Gharibyan}
\address{Yerevan~Physics~Institute,~Armenia } 

\begin{abstract}
Current constraints on photon velocity variability are summarized and displayed
in terms of an energy dependent vacuum refraction index. It is shown that
the energy-momentum balance of high energy Compton scattering  is very sensitive 
to the outgoing photon speed. A missing energy observation in HERA Compton 
polarimeter data indicates that photons with 12.7 GeV energy are moving 
faster than light by 5.1(1.4)mm/s. 
An asymmetry spectrum measured by the SLC longitudinal polarimeter implies 
however an effect which is 42 times smaller, although the interpretation of the            
data is less clear here.
\end{abstract}

\end{frontmatter}

\section{ Theoretical Models}
According to relativistic kinematics a photon velocity   
in vacuum $c_\gamma$ does not depend on its energy $\omega$, 
while a possible dependency is constrained by the current photon mass  
limit \hbox{$m_\gamma\!<\!10^{-16}~eV$~\cite{Groom:2000in}} as 
\hbox{$1-c_{\gamma}(\omega)/c\leq 10^{-32}\omega^{-2}~eV^2$,} where $c$ is a 
massless particle vacuum speed.
However, the laboratory or stellar vacuum always contains background 
fields (matter) and  quantum interactions can slow down
or speed up photon propagation. Tiny changes of the photon velocity have 
been predicted~\cite{Latorre:1995cv},\cite{Dittrich:1998fy}
for such non-trivial, polarized vacua modified by 
electromagnetic or gravitational fields, temperature or boundary conditions 
within the perturbative quantum electrodynamics which allows
to derive inverse relative velocities (vacuum refraction indices 
$n\!=\!c/c_\gamma$) mainly for low energy $\omega\ll m$ ($m$ is the electron 
mass) photons~\cite{Scharnhorst:1998ix}. Even in the absence of background 
fields vacuum quantum fluctuations can influence light propagation as  
pointed out for the gravitational vacuum by recent developments in quantum 
gravity theory~\cite{QGrav}. Changes of photon speed 
are expected to be significant at photon energies close to the Planck mass 
$\approx10^{19}~GeV$ decreasing with lower energies. 
Hypothetical Lorentz symmetry deformations considered for explaining 
the observed ultrahigh energy cosmic rays above the GZK cutoff
(and possibly neutrino oscillations)~\cite{Coleman:1999ti}
may also introduce an energy dependent photon speed~\cite{Amelino-Camelia:2000zs}. 
\section{ Experimental Limits}
Magnitudes of these predicted effects are  small and 
though may exceed by many orders the constraints imposed by the photon 
mass, all experimental tests so far show that different energy photons 
in vacuum move at the same velocity (light vacuum speed $c$) within the 
constraints displayed on fig.1 (use of vacuum refraction index $n(\omega)$ 
instead of photon velocity is convenient to distinguish between photon mass
and vacuum properties).
\begin{figure}[h]
\centerline{\epsfig{file=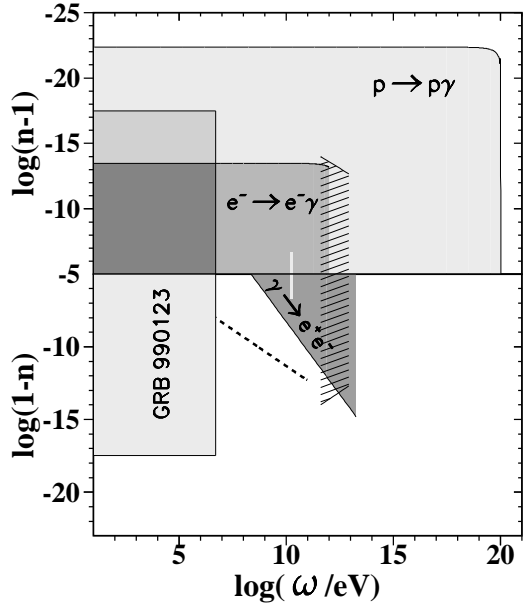,height=8cm, angle=0}}  
\caption{\label{fig1}
Experimental constraints on the vacuum refraction index.}
\end{figure}

The most stringent limits are coming from the detection of highest energy
proton and $\gamma$ cosmic particles as first noted in~\cite{Coleman:1997xq}, 
since in a dispersive vacuum they would quickly decay by vacuum Cherenkov 
radiation \hbox{$p\rightarrow p\gamma$ ($n>1$)} and pair creation 
\hbox{$\gamma\rightarrow e^+e^-$ ($n<1$)}. 
These processes are kinematically forbidden in case
\begin{equation}
  n-1<\frac{M^2}{2E^2-2\omega E-M^2};\hspace{10mm}
  1-n<\frac{2m^2}{\omega^2}   
 \label{eqlim} 
\end{equation}
for  Cherenkov radiation and pair creation respectively with $M,E$ 
the proton mass and energy.
Excluded areas in fig.1 correspond to a highest detected proton energy 
of $E=10^{20}~eV$~\cite{Wdowczyk:1989ub} and to a cosmic photon spectrum up to 
$\omega_{max}=22~TeV$~\cite{Konopelko:1996er}. Also shown is a limit inferred 
from the highest observed electron energy of $2~TeV$~\cite{Kobayashi:he}.
Other areas are excluded by experiments utilizing direct time of flight techniques
sensitive to $\left|n-1\right|\approx \Delta tc/D$, where $\Delta t$ is a time 
difference between arrivals of simultaneously emitted photons with different 
energy and $D$ is a distance 
to the source. While laboratory experiments are limited by
time resolutions of typically a few $psec$ and  distances of a few $km$ (an early 
SLAC result~\cite{Guiragossian:1975wp} $\left|n-1\right|< 2\cdot10^{-7}$  is 
shown on fig.1 by a narrow white bar at \hbox{$15~GeV\!<\!\omega\!<\! 20~GeV$)} the 
astrophysical observations could do much
better owing to huge distances to the source. In ref.~\cite{Schaefer:1998zg} one 
can find limits on light speed variations in wide energy ranges based on 
different astrophysical events; these limits suffer, however, from very uncertain 
distance scales.
Meanwhile an observed spectacular gamma ray burst GRB990123~\cite{Akerlof:1999we}
followed by an optical counterpart detected within \hbox{$\Delta t\!=\!22sec$}, 
with a distance \hbox{$z\!=\!1.6$},  could establish a constraint 
\hbox{$\left|n-1\right|\!<\!3\cdot10^{-18}$}  for 
\hbox{$2~eV\!<\!\omega\!<\!5~MeV$,}
which is anyhow the order of constraints quoted in ref.~\cite{Schaefer:1998zg}.  
Photons with highest observed energies  
\hbox{$0.35~TeV\!<\!\omega\!<\!10~TeV$} 
from a well defined  active galaxy source (Markarian 421) put constraints 
$\left|n-1\right|<2.5\cdot10^{-17}\omega$ \cite{Biller:1999hg} (hatched area 
in fig.1).
\section{ Compton scattering in dispersive vacuum}
Apart from the discussed threshold effects for vacuum Cherenkov and pair 
creation, the dispersive vacuum will modify the kinematics 
of other processes involving free photons according to the dispersion relation 
\hbox{$k^2\!=\!\omega^2(1-n^2)$.} However, the tiny 
refraction imposed by such vacuum becomes observable only
at high energies with corresponding small angles. 
When the photon (four-momentum $k$) interacts with 
a particle (four-momentum $P$) the vacuum index will contribute 
to the convolution $Pk$ as  
\begin{equation}
Pk\approx \frac{\mathcal E\omega}{2}\Biggl(\frac{1}{\gamma^2}+
\theta^2+2(1-n)\Biggr)
\label{eqpk}
\end{equation}
where $\mathcal E,\gamma \gg$1 are energy, Lorentz-factor
of the particle, and  $\theta\ll 1$ is the angle between the photon 
and the particle.
Thus, such processes in general could detect a relative 
photon speed variation, at given energy $\omega$, as small as the order of
$1/2\gamma^2$. 

Below we concentrate on photon scattering off an ultrarelativistic electron
and apply (\ref{eqpk}) in energy-momentum conservation to get sensitivity of  
the high energy Compton process to the vacuum refraction index. 
If $\omega_0, \theta_0, \omega, \theta$ designate energy and angle of the 
incident and scattered photons, for $\omega_0\ll\omega\gg m$ we have
\begin{equation}
   n-1=\frac{1}{2\gamma^2 }\Biggl[
   1+\theta^2\gamma^2-x\Biggl(\frac{\mathcal E}{\omega}-1\Biggr)\Biggr]
\label{eqnm1}
\end{equation}
where $\gamma,\mathcal{E}$ are the Lorentz-factor and energy of the initial 
electron,
\[x\equiv \frac{4\gamma \omega_0\sin^2{(\theta_0/2)}}{m}\],  
and $n$ is the index for the direction $\theta$ and energy $\omega$.  
In a case of  laser Compton scattering on accelerator electrons
the initial states ($x,\gamma$) are known to high degree of precision
(typically to $0.01\% $)
 which allows to gain information about $n$ from 
each event measuring  the $\omega$ and $\theta$ (or the energy 
and angle of the scattered electron $\mathcal{E}^\prime, ~\theta^\prime$, since 
\hbox{$\omega\!=\!{\mathcal E}-{\mathcal E}^\prime,
~\theta\!=\!\theta^\prime\mathcal{E}^\prime/\omega$}).
Alternatively one could detect only the Compton edge i.e. maximal(minimal) 
energy of the scattered photons(electrons) \hbox{$\omega_m\!\equiv\!\omega$} at 
\hbox{$\theta\!=\!0$} 
\hbox{(${\mathcal E}_m^\prime\!\equiv\!{\mathcal E}-\omega_m$)}
to measure $n(\omega_m)$  down to values of
\begin{equation}
   |n-1|\leq\frac{2\omega_0}{\omega_m }\frac{\Delta\omega_m}{\omega_m}
\label{eqnmm}    
\end{equation}
which follows from (\ref{eqnm1}) if $\omega_m$ is measured with relative
uncertainty $\Delta\omega_m/\omega_m$.
A dotted line in fig.1 shows the potential of  
laboratory Compton scattering in limiting $n$ according  to (\ref{eqnmm}) 
for optical 
lasers ($\omega_0 \approx 2~eV$) with a modest precision 
$\Delta\omega_m/\omega_m=1\%$ up to a photon energy of $100~GeV$.  

The laser scattering is particularly attractive to test vacuum 
birefringence since the highest energy scattered photons preserve 
the laser polarization~\cite{Babusci:ji} which is easy to change. 
Flipping the laser linear polarization one could measure $n_{\perp}$, 
$n_\parallel$ components for multi-GeV photons by detecting the Compton
edge dependence on $\perp$, $\parallel$ polarization states (current
bounds on the vacuum birefringence~\cite{Kostelecky:2001mb} are set
by polarimetry of (near)optical photons coming from distant 
astronomical sources). 

In ref.~\cite{Gurzadian:at} it has been proposed to test different quantities
related to photon velocity by high energy Compton process measuring 
simultaneously the scattered photon and electron energies. However, this 
set of measurements is not sensitive to the photon speed which is 
accessible only from the photon energy and momentum combined information.
To measure the photon momentum one has to register scattering angle
of the photon or electron relying for the latter case on energy-momentum 
conservation. It is possible to indirectly register zero scattering angle
of the photon by detecting the Compton edge as it pointed out above and
only then an energy measurement alone is sufficient to obtain information 
about the photon speed. 
 Another distinguished  kinematic point in the Compton process, where 
 circularly polarized photons interact with longitudinally polarized 
 electrons, is the energy asymmetry (between spin 1/2 and 3/2 states) zero 
 crossing which occurs at the maximal scattering angle of the electron and 
 therefore at a fixed photon momentum. Thus, the corresponding energy 
 $\omega_{A=0}$ of the scattered photon gives a measure of the photon speed
 \begin{equation}
   n-1=\frac{1}{2\gamma^2 }\Biggl[
   1-\frac{x}{2}\Biggl(\frac{\mathcal E}{\omega_{A=0}}-1\Biggr)\Biggr]
 \label{eqnm11}  
\end{equation}
 This is a reduced form of eqn.(\ref{eqnm1}) where angle detection is 
 replaced by an energy measurement at the expense of dealing with polarized beams
 and is useful because most of the laboratory Compton devices are working as 
 polarimeters. 

Derived relations  allow to extract the refraction index and associated
photon speed from existing polarimetric data.

\section{HERA polarimeter spectra analysis} 

 Consider photon spectra (fig.2) from ref.~\cite{Barber:1992fc}
measured by the HERA Compton polarimeter. 
\begin{figure}[h]
\centerline{\epsfig{file=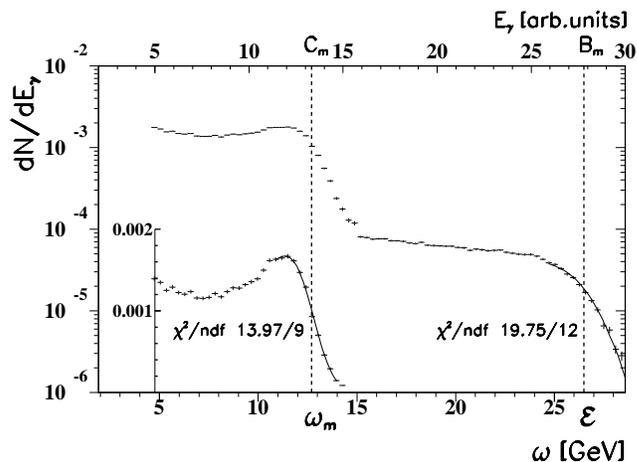,height=6cm, angle=0}} 
\caption{\label{fig2}
HERA polarimeter Compton events on top of background Bremsstrahlung
and background subtracted Compton spectrum (inset) with fit results.  
Upper scale: original energy calibration using nominal Compton edge
(GeV replaced by arb.units).
Lower scale: recalibration using Bremsstrahlung edge.}
\end{figure}
The spectra were obtained by directing a CW $514.5~nm$ laser 
light against the HERA transversely polarized, $26.5~GeV$ electron 
beam with a vertical crossing angle of $3.1~mrad$ and detecting produced 
high energy $\gamma$-quanta with a sampling calorimeter.  
The whole detection scheme is designed for measurement of an up-down spatial
asymmetry of the $\gamma$-quanta which is introduced by a flip of the laser 
light helicity
and is proportional to the electron beam polarization while the energy
measurement is auxiliary and serves as a mean to enhance the asymmetry by proper
energy cuts. 
We are going to extract Compton $\gamma$'s maximal energy from the spectra
and estimate the refraction index via eqn.(\ref{eqnm1}) at \hbox{$\theta\!=\!0$}.
Hence, following ~\cite{Barber:1992fc} and~\cite{Lomperski:1993aw}, we 
concentrate on details of the experimental setup important for energy
measurement only, ignoring all features related to polarization.

The scattered Compton photons originate from an interaction region (IR) 
about $50~cm$ long, defined by the crossing angle and size of the electron and 
laser beams. Bending magnets downstream of the IR separate the electron 
and $\gamma$ beams and the photons leave the vacuum pipe through a 
$0.5~mm$ thick aluminum window to pass $39~m$ of air before entering the 
calorimeter which is installed $65~m$ away from the IR. 

Collimators placed at a distance of $47~m$ from the IR, define an aperture of 
$\pm 0.37 mrad$ the same as angular size of the calorimeter as seen from 
the IR. The aperture is 15 times larger than the largest (horizontal) angular 
spread of electrons at the IR and 40 times larger than the characteristic 
radiation angle $1/\gamma$ so the acceptance inefficiency can be ignored. 
The collimators are followed by magnets to sweep out charged background.
    
The calorimeter consists of 12 layers of $6.2~mm$ thick tungsten and $2.6~mm$
thick 
scintillator plates surrounded by 4 wavelength shifters attached to 4
photomultipliers. PMT signals from single photons are integrated within $100~ns$
gate then digitized with 12 bit ADCs and gains of the PMTs are adjusted to 
about $15~MeV$ per ADC channel. 
A fast DAQ handles the signals and operates without dead time up to an average 
data rate of  $100~kHz$.  
The detector performance has been simulated with EGS4 Monte-Carlo program and 
tested using DESY and CERN test beams. Measured energy resolution of 
$24\%~GeV^{1/2}$, spatial non-uniformity of $\pm 1\%$ and nonlinearity of $2\%$ at 
$20~GeV$ are reported to be in agreement with the simulations.

Apart from the laser light, the electron beam also interacts with residual gas,
thermal photons and bending magnetic field in the beam pipe producing respectively
Bremsstrahlung, scattered blackbody and synchrotron radiation reaching the 
calorimeter. To measure this background the laser beam is blocked for $20~sec$
of each $1~min$ measurement cycle (light on/off is 40/20~sec). The procedure 
allows to eliminate the background by a simple subtraction of time normalized 
light-off spectrum from the light-on spectrum. Exact on/off durations are 
counted by DAQ clocks.

 At the time of the measurements an electron beam current of 
$0.32~mA$ and a laser power of $10~W$ provide $1~kHz$ rate above an energy 
threshold of $1.75~GeV$ while the background rate was $0.15~kHz$.
With such high threshold only the Bremsstrahlung contributes to background since 
the scattered blackbody radiation maximal energy is $0.73~GeV$ and the synchrotron
radiation is absorbed in the first tungsten plate of the calorimeter. 

Putting the laser photon, HERA electron energy and the crossing angle  
($\omega_0\!=\!2.41~eV, ~{\mathcal E}\!=\!26.5~GeV,~\theta_0\!=\!\pi+3.1~mrad$)
into the definition of the kinematic parameter $x$, we get $x\!=\!0.9783$. 
The precision of the parameter is limited by the electron beam energy uncertainty 
$\sigma (\mathcal E)/\mathcal E\approx 10^{-4}$. Errors of the other constituents
$\sigma (\omega_0)/\omega_0\approx 10^{-5}$,
$\sigma (m)/m \approx 3\cdot10^{-7}$,
$\Delta (\theta_0)\!\approx\!2~mrad\Rightarrow$ 
$\Delta\sin^2{(\theta_0/2)}\approx 3\cdot10^{-6}$ contribute negligibly.
  
 To measure the ratio ${\mathcal E}/\omega_m$  (the only unknown
in the right part of (\ref{eqnm1}) at \hbox{$\theta\!=\!0$)} we can utilize
the Bremsstrahlung spectrum (\cite{Barber:1992fc}, fig.18) 
which helps to cancel the absolute energy calibration of the calorimeter since
\begin{equation}
\frac{\mathcal E}{\omega_m}=\frac{\alpha B_m + m}{\alpha C_m}
=\frac{B_m}{C_m}+O(4\cdot10^{-5})
\label{eqrat}
\end{equation}
where $\alpha$ is a calibration constant and $B_m$, $C_m$ are the 
Bremsstrahlung and Compton edges derived from the measured spectra 
in arbitrary units. It is easy to verify that influence of
the term (\ref{eqpk}) to the Bremsstrahlung maximal energy is negligible
i.e. a non-zero $|n-1|$ shifts only the Compton edge. 
  
A spectrum measured via calorimetry is conventionally described by a 
function
\begin{equation}
F(E_\gamma)=N\int^{E_m}_{0} 
\frac{d\Sigma}{d\omega}\frac{1}{\sqrt{\omega}} 
\exp\Biggl({\frac{-(\omega-E_\gamma)^2}
{2\sigma_0^2 \omega}\Biggr) d\omega}
\label{eqfold}
\end{equation}
where a parent energy distribution ${d\Sigma}/{d\omega}$ incident on
the detector is folded with a response function which is 
a gaussian with energy dependent width equal to the calorimeter energy 
resolution (in our case {$\sigma_0\!=\!0.24~GeV^{1/2}$}), $N$ is a normalization
constant and $E_m$ is the cutoff energy of the parent distribution.   

The original energy calibration is made to match the nominal Compton edge 
$\mathcal E x/(1+x)\!=\!13.10~GeV$  (fig.2, upper scale) by applying a 
differentiation deconvolution method to find the cutoff energy. This method 
unfolds the spectrum by numerical differentiation to reveal a nearly gaussian 
peak (inverted) within the spectrum fall-off range and assigns the peak position 
to the cutoff value. The main drawback of this method comes from ignorance of the 
parent 
distribution which results in a shifted answer in case of non-flat distributions 
as follows from (\ref{eqfold}). 
Therefore, to extract the $B_m$, $C_m$ values from the 
spectra we have used a more precise approach (fitting via (\ref{eqfold})) and have
applied the differentiation method only to find the fit ranges around 
end points where the differentiated spectra peak, since outside of 
these ranges the spectra contain no information about the cutoff  energies. 
Such localization also helps to avoid possible bias of the fit results caused by 
physical effects affecting the spectra and not entering in the function 
$F(E_\gamma)$.
Dominating among these effects are photon conversions between the interaction 
point and the calorimeter, detector nonlinearity and spatial non-uniform 
response. 
These effects change the shape of the 
spectra in a way the function (\ref{eqfold}) is not able to describe adequately 
over the full energy range which is expressed also in 
ref.~\cite{Barber:1992fc} and is noticeable for original fits shown on fig.18 and 
fig.21 of the same ref.

Fitting the function $F(E_\gamma)$ to the background spectrum with the beam-gas
Bremsstrahlung cross-section~\cite{Andruszkow:1992rz} as the parent distribution 
and 2 free, variable parameters $E_m$, $N$,  we get 
{$E_m\!=\!27.799\pm 0.047\!=\!B_m$, } (see fig.2). The fit range is predefined by
numerical differentiation of the spectrum as discussed above.
From a similar fit to the background subtracted Compton 
spectrum (from fig.21 of ref.~\cite{Barber:1992fc}) by $F(E_\gamma)$
with the Compton cross-section~\cite{Ginzburg:1982yr} as parent distribution, we 
find \hbox{$E_m\!=\!13.322\pm0.010\!=\!C_m$} (fig.2 inset).
According to the derived numbers $B_m$,$C_m$ and relations
 (\ref{eqrat}),(\ref{eqnm1}) 
we have the Compton edge at \hbox{$\omega_m\!=\!12.70\!\pm\! 0.02~GeV$,} 
well below from the nominal {$\omega_m(n=1)=13.10~GeV$} value and a 
vacuum index for the $12.7~GeV$ photons
{$n\!=\!1-(1.17\!\pm\!0.07)10^{-11}$} which is responsible for such 
reduction. 

Now we return to the above mentioned systematic effects to estimate their
possible influence on the obtained cutoff energies. 
The non-evacuated path of $\gamma$ beam line serves as an extended target to 
convert them into $e^+ e^-$ pairs, subject to continuous energy loss and multiple 
scattering before registration by the calorimeter. 
This modifies the spectra by enhancing lower energy parts without 
affecting the highest detected energies from non-converted $\gamma$-quanta.  
The most significant instrumental source affecting the result is the detector 
non-linear response $E\alpha^{-1}(1+fE)$ under a given energy $E$ with 
\hbox{$f\!=\!-0.001~GeV^{-1}$}
from the quoted nonlinearity of $2\%$ at $E=20~GeV$
($f<0$ corresponds to a conventional calorimetric nonlinearity arising from 
shower leakage). This brings the ratio (\ref{eqrat}) to
\begin{equation}
\frac{B_m}{C_m}\approx\frac{\mathcal E}{\omega_m}(1+f({\mathcal E}-\omega_m))
\end{equation}
with a corresponding correction of $0.52\cdot 10^{-11}$ for $n$
and half of that value as the correction error.

Another possible source of the edges mismatch would arise if the 
Bremsstrahlung and Compton beams incident on calorimeter are separated in 
space. Propagating the quoted spatial non-uniformity 
of the calorimeter $\pm 1\%$ ({$B_m/C_m \rightarrow B_m/C_m (1\pm 0.01)$}) to the
value of $n$ we finally have 
$$n=1-(1.69\pm0.07\pm0.38\pm0.26)10^{-11}$$
with statistical and systematic non-uniformity, nonlinearity errors displayed 
separately. 

For completeness of the analysis, we discuss a few additional systematic sources
which have no significant effect on the energy distributions. 
First is the ADC electronic pedestal with a width equal to $\approx\!30~MeV$ and 
a measured systematic shift of the mean of $\pm 4~MeV$. This is neglected, 
since the pedestal spread is incorporated into the energy resolution while the 
shift is 
less than the result's smallest, statistical error by almost one order of 
magnitude.  
Next is an emission of multiple photons by a single electron bunch resulting
in enhanced maximal detected energies due to a pile-up in the calorimeter. 
Using a Poisson distribution and evaluating the quoted single photon emission 
probability $p_C (1)=0.02$ at the given Compton rate, one readily has probabilities 
for 2 photon emission 
$p_C (2)=2.2\cdot10^{-4}$ 
and $p_B (2)=9.9\cdot10^{-6}$ for Compton and Bremsstrahlung 
respectively. The latter number is too small to cause any considerable shift
of $B_m$, since the whole Bremsstrahlung spectrum contains less than a few 
pile-up events. Concerning the Compton edge, correcting for pile-up 
would only aggravate the observed energy reduction. 
The same is true also for non-linear Compton scattering events where an 
electron emits two or more photons at once.

\section{SLC polarimeter asymmetry analysis}

In ref.~\cite{Shapiro:1993gd} one can find a Compton asymmetry 
(fig.3)  measured by the SLC polarimeter where high power laser pulses of $532~nm$ 
circular light interact with longitudinally polarized bunches of $45.6~GeV$ 
electrons under a crossing angle of $10~mrad$ and recoil electrons are registered
by an array of Cherenkov counters installed downstream of two momentum analyzing 
bending magnets. Each channel of the detector integrates multiple electrons per 
pulse, within a certain energy range according to its position in the array.
Following a detailed description of the polarimeter setup in~\cite{King:1994yt}
one infers that energies detected in the $N$-th channel are constrained by 
\begin{equation}
{\mathcal E}'_{min(max)}=C_{0}(S_N+(-)D/2+S-S_{c})^{-1}
\label{eqhorz}
\end{equation}
where $C_{0}\!=\!296.45~GeV\cdot cm$, $S\!=\!10.58cm$, $S_N\!=\!N~cm$, $D\!=\!1~cm$ 
which is the channel size and $S_c$ is the Compton kinematic endpoint distance 
from the channel 7 inner edge which also depends on the initial electron beam 
position relative to the detector. 
\begin{figure}
\centerline{\epsfig{file=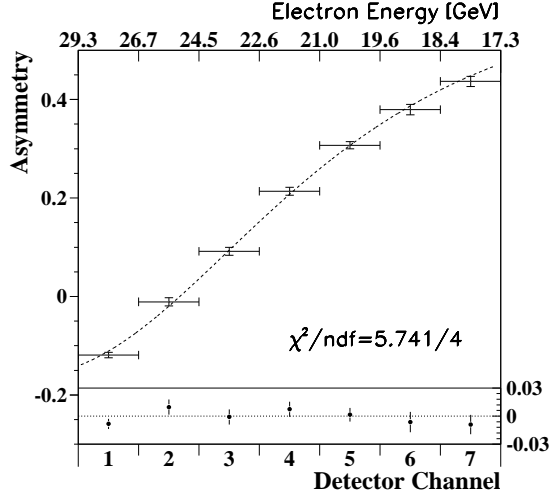,height=6.5cm, angle=0}} 
\caption{\label{fig3}
SLC polarimeter asymmetry (lower scale) with fit results (upper scale).
The dotted line shows the parent distribution 
$d\Sigma_{\lambda}/d\Sigma_c$.
The lower part displays the fit residuals (right scale).}
\end{figure}
Information about the photon speed is encoded into the relation of the Compton 
maximal and asymmetry zero crossing energies according to eqn.(\ref{eqnm11}). 
A coarse granularity of the detector (binning in fig.3), however, makes it 
difficult to apply this simple kinematic method. Instead one can utilize 
dynamic features of the Compton scattering in the case of $n\neq 1$. 
Using an invariant representation of the Compton process in 
ref.~\cite{Ginzburg:1982yr},
for longitudinal polarization of the incident electron beam one can write 
the cross-section as 
\begin{equation}
\frac{d\Sigma_c}{dy}+\lambda \frac{d\Sigma_\lambda}{dy}=
\frac{\pi r_e^2}{x}\Biggl( \frac{1}{1-y}+1-y-4r(1-r)
+\lambda u
\Biggr) 
\label{eqnxs}
\end{equation}
where $r_e$ is the classical electron radius, $\lambda$ is the electron beam and 
circular light polarizations product, $r\!=\!y/(x-xy)$, $u\!=\!rx(1-2r)(2-y)$, 
$y\!=\!1-Pk/Pk_0$ with $k_0$ being the photon's initial four-momentum,
and $x=2Pk_0 /m^2$, which is the kinematic parameter defined above.
  
To introduce a refraction index into the cross-section, we modify $Pk$
entering in $y$ according to (\ref{eqpk})
  and scale cross-section (\ref{eqnxs}) by a factor of  
$$\Bigl(n^2+n\omega\frac{dn}{d\omega}\Bigr)^{-1}$$
 which accounts for a change of the 
delta function $\delta (\omega^2 -\mathbf{k}^2)$ to 
$\delta (n^2 \omega^2 -\mathbf{k}^2)$ in the phase space of the outgoing photon.
In addition we use (\ref{eqnm1}) and energy conservation to eliminate $\theta$ 
and express the cross-section in terms of $\mathcal E '$. 
  
The asymmetry $A_N$ measured in a given detector channel $N$ is a product of
$\lambda$ and an analyzing power $I_\lambda/I_c$ 
($A_N=\lambda\cdot I_\lambda/I_c$) where
\begin{eqnarray}
I_{\lambda(c)}=\int^{{\mathcal E}'_{max}}_{{\mathcal E}'_{min}} 
{\mathcal E}'\frac{d\Sigma_{\lambda(c)}}{d{\mathcal E}'}d{\mathcal E}'
\label{asym}
\end{eqnarray}
with ${\mathcal E}'$ being the scattered electron energy
limited by the channel's energy acceptance 
${\mathcal E}'_{min}$, ${\mathcal E}'_{max}$. 

It follows from (\ref{eqhorz})-(\ref{asym}) that in the case of $n\!=\!1$,
 the parameters
$S_c$ and $\lambda$ establish horizontal and vertical scales respectively
(energy and asymmetry) in fig.3.
However, these variables alone are not sufficient for a satisfactory description
 of the asymmetry distribution as indicated by a least squares fit performed
with only two free parameters $S_c$ and $\lambda$. 
Ref.~\cite{King:1994yt} also reports about interchannel inconsistencies which
dictate the choice and use of only one channel (number 7) for the polarization
measurement.

To extract the photon speed we add one more free parameter 
$\psi\equiv 2\gamma^2 (n-1)$ and use the polarized Compton cross-section modified 
by dispersion, assuming a constant refraction index across the entire 
energy range of the measured asymmetry.
Now the $\chi^2$ minimization converges with 
$\lambda=0.628\pm 0.009$, $S_c =0.970\pm 0.037$ and
$\psi=-(6.49 \pm 0.08)10^{-3}$ (fig.3), which yields
 $$n=1-(4.07\pm  0.05)10^{-13}$$
for photons in the energy range $16.3~GeV<\omega<28.3~GeV$. 

An influence of the detector response on the asymmetry is quoted 
in~\cite{King:1994yt}
to be about 1\%, which is much smaller than statistical fluctuations
and we ignore it.
Assuming perfect circular polarization of the laser light,
$\lambda$ equals the electron beam polarization, which is measured to be
$0.612\pm 0.014$ from the channel 7 asymmetry. Both numbers agree within 
statistical and declared 1.41\% systematic~\cite{King:1994yt} errors, 
and at the same time the $n\neq 1$ hypothesis allowed 
the asymmetry spectrum to be fitted successfully. 

Although the obtained result is more precise compared to the HERA observation,
it is less reliable because of the multi-electron generation-detection scheme
and a theoretical uncertainty.  
The multi-particle mode, in general, poses difficulties to separate and treat the 
systematics and it also forced us to abandon the clear kinematic approach utilized 
in the case of the HERA polarimeter, while
the method applied for modification of the Compton cross-section is
somewhat heuristic and may introduce theoretical errors.  

\section{Discussion}
The observed value of the index, obtained from one sample of the HERA 
polarimeter data, is statistically significant and does not contradict 
any previous experimental result (fig.1).
It is below unity testifying that $12.7~GeV$ energy photons are moving 
faster than light (by $c(1-n)=5.07\pm 1.41~mm/s$). 
However a SLAC experiment shows that for photons of energy $16.3-28.3~GeV$, the 
departure from the speed of light is at most $0.122\pm0.0015~mm/s$.

Although the sign of the effect alone may be favorable for some theories
discussed in sec.1, the detected magnitude is too large 
to be associated with polarized electromagnetic or gravitational 
vacuum. So, the outcome is unexpected, especially in view
of the sharper limits for surrounding energies (see fig.1) and it is 
interesting to see whether the result can stand an examination by 
dedicated measurements and/or rigorous analysis of other pieces 
of data. 
\ack
The author would like to thank A.~T.~Margarian for useful 
discussions and to M.~Lomperski for providing 
measurement details and pointing to some sources of systematic uncertainties 
of the HERA polarimeter.

\end{document}